\documentclass[twoside,twocolumn,11pt]{article} 
\DeclareUnicodeCharacter{202F}{FIX ME!!!!} 
\usepackage{blindtext} 
\usepackage[sc]{mathpazo} 
\usepackage[T1]{fontenc} 
\usepackage{microtype} 
\usepackage{csquotes}
\usepackage[english]{babel} 
\usepackage{graphicx}
\usepackage{float}
\usepackage[linguistics]{forest}
\usepackage{wrapfig}
\usepackage{amsmath,amssymb}
\usepackage{graphicx}
\usepackage{subcaption}

\usepackage[hmarginratio=1:1,top=32mm,columnsep=20pt,margin=0.6in]{geometry} 
\usepackage[labelfont=bf,textfont=it]{caption} 
\usepackage{lettrine} 
\usepackage{enumitem} 
\setlist[itemize]{noitemsep} 
\usepackage{abstract} 
\usepackage{titlesec} 
\renewcommand\thesection{\Roman{section}} 
\renewcommand\thesubsection{\roman{subsection}} 
\titleformat{\section}[block]{\large\scshape\centering}{\thesection.}{1em}{} 
\titleformat{\subsection}[block]{\large}{\thesubsection.}{1em}{} 
\usepackage{fancyhdr} 
\usepackage{titling} 
\usepackage{cite}
\usepackage{hyperref} 
\pretitle{\begin{center}\Huge\bfseries} 
\posttitle{\end{center}} 
\title{Electrical Steel: An investigation into the
brittleness of the Fe-Si alloy} 
\author{%
\textsc{Arthur Adriaens} \\[1ex] 
\normalsize Ghent University \\ 
\normalsize \href{mailto:arthur.adriaens@ugent.be}{arthur.adriaens@ugent.be} 
\and 
\textsc{Cedric Ooms} \\[1ex] 
\normalsize University of Antwerp \\ 
\normalsize \href{mailto:cedric.ooms@student.uantwerpen.be}{cedric.ooms@student.uantwerpen.be} 
\and 
\textsc{Corentin Mergny} \\[1ex] 
\normalsize Free University of Brussels \\ 
\normalsize \href{mailto:corentin.chris.m.mergny@vub.be}{corentin.chris.m.mergny@vub.be} 
}

\date{\today} 
\renewcommand{\maketitlehookd}{%
\begin{abstract}
\noindent 
In this paper, we'll hunt for ordered crystal structures that may be
stable, and that would explain the brittleness of electrical steel. I.e we wish to find crystals with negative formation energy, if it turns out that these structures are very anisotropic that would be an indication of brittleness (as a ductile material will generally be homogeneous).
We'll do this by exploiting the advantage that ab initio simulations (done using Quantum
ESPRESSO (opEn-Source Package for Research in Electronic Structure,
Simulation, and Optimization) \cite{Giannozzi_2017} \cite{Giannozzi_2009} \cite{exoscaleqe}) give full control over defining
the crystal: We'll tell exactly where we want to have every atom, and with DFT we'll calculate the  internal energy that corresponds to such a
crystal. The whole project is open source on \href{https://github.com/arthuradriaens-code/Fe-Si}{github} and as we haven't found the answer yet, everyone who is interested is welcome to contribute.
\end{abstract}
}

\begin{document}

\maketitle


\section{Introduction}
\lettrine{E}{ectric} applications such as motors, transformers or generators all have a magnetic material in the core of their electromagnetic coil. In most cases this is a so-called electrical steel: an Fe-Si alloy with about 3 wt.\% Si.
It is known since
decades that using a steel with 6.5 wt.\% Si would be very advantageous\cite{6.5wt} over the steel we're now using. Such a steel would
reduce energy losses in the application due to heat, which would make it possible to build electric machines lighter and more
energy efficient. Estimates come up with a saving of $\approx 12$ billion euros worth of electricity every
year.
So why aren't we using that ideal electrical steel? In contrast to 3 wt.\% Si, the 6.5 wt.\% Si steel is
brittle: you can’t press or roll or otherwise form it into the size and shape needed to build the electric
apparatus. It would just break apart when trying to do so. Hence finding an electrical steel with 6.5 wt.\% Si
that is not brittle is quite the holy grail in electrical steel research.
There is a hypothesis about why the brittleness appears. Crystals with long range order are usually
more brittle than crystals in which the atoms are more disordered. It is assumed that when
increasing the silicon content, there is a stronger tendency for the atoms to develop short-range
order. In this paper, we'll investigate that hypothesis by constructing various unit cells.

\section{Convergence testing}
We'll be using the SPSS pseudopotentials\cite{doi:11.1126/science.aad3000} \cite{Prandini2018} for all of our calculations.
Two stable crystals out of which our (later on) proposed unit
cells could be made out of (or deteriorate to) are bcc iron and 
DO3-Fe$_3$Si, it's thus necessary to first determine all the
used parameters with respect to these crystals.
\subsection{bcc-Fe}
\begin{wrapfigure}{R}{0.1\textwidth}
  \begin{center}
	  \includegraphics[width=0.1\textwidth]{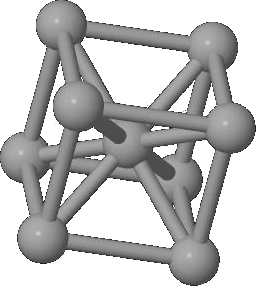}
  \end{center}
\end{wrapfigure}
For bcc-Fe iron we use Patrick M. Woodward et.al's cif file. We're also using a pseudopotential generated using "atomic" code by A.
Dal Corso  v.5.0.99 svn rev. 10869 with the minimum cutoff for wavefunctions
being the suggested 64. Ry and the minimum cutoff for charge density the
suggested 782. Ry. We use the hydrostatic pressure to test the convergence, for different k-meshes the results are shown in table \ref{table:fek-mesh}.
\begin{table}
\begin{tabular}{c|c|c}
	& k mesh & Hydrostatic Pressure (kbar)\\
	\hline
	1&(1,1,1)&291.45\\
	2&(3,3,3)&76.49\\
	3&(5,5,5)&-99.45\\
	4&(7,7,7)&-71.98\\
	5&(9,9,9)&-109.31\\
	6&(10,10,10)&-98.83\\
	7&(11,11,11)&-80.23\\
	8&(13,13,13)&-92.67\\
	9&(15,15,15)&-84.76\\
\end{tabular}
\caption{Fe k-mesh convergence}
\label{table:fek-mesh}
\end{table}\\
We find it has sufficiently converged at (10,10,10) so that's the k-mesh we'll
be using. Now keeping the cutoff charge density $\approx 12$ times the cutoff
for wavefunctions, we'll vary this wavefunction cutoff as shown in table \ref{table:feecutwfc}
\begin{table}
\begin{tabular}{c|c|c}
	&ecutwfc&Hydrostatic Pressure (kbar)\\
	\hline
	1&14&-15244\\
	2&24&-1727\\
	3&34&-962\\
	4&44&-457\\
	5&54&-95\\
	6&64&-98\\
\end{tabular}
\caption{Fe ecutwfc convergence}
\label{table:feecutwfc}
\end{table}\\
Here we see quite a good convergence at ecutwfc = 54, lastly we'll take ecutwfc=54 and vary the multiplicity of the charge density as shown in table \ref{table:fefactor}.
\begin{table}
\begin{tabular}{c|c|c|c}
	&factor&ecutrho&Hydrostatic Pressure (kbar)\\
	\hline
	1&2&108&939\\
	2&4&216&0.81\\
	3&6&324&-69\\
	4&8&432&-96\\
	5&10&540&-92\\
	6&12&648&-95\\
\end{tabular}
\caption{Fe factor convergence}
\label{table:fefactor}
\end{table}\\
We already see convergence at a factor 8. Our final values are thus a k-mesh of (10,10,10), ecutwf at 54 and ecutrho at 432.
\subsection{DO3-Fe3Si}
\begin{wrapfigure}{L}{0.1\textwidth}
  \begin{center}
	  \includegraphics[width=0.1\textwidth]{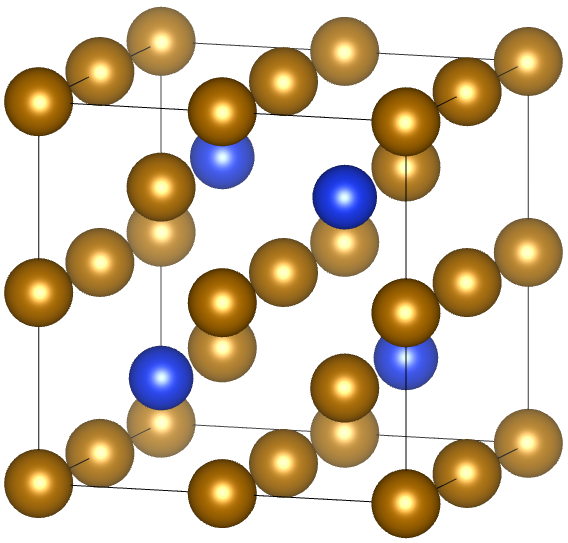}
  \end{center}
\end{wrapfigure}
DO3-Fe$_3$Si is the name of the crystal with a unit cell built from 8 bcc-Fe
unit cells stacked in a 2x2x2 way, to make a cube with twice the edge length of
a normal bcc-Fe unit cell. Herein every other unit cell's middle iron nucleus
is replaced with a Si nucleus, such that these 4 atoms form a tetrahedron. As shown 
on the inline figure\footnote{cif file by Farquhar M. C. M.,
Lipson H. and Weill A. R., with COD index 9015110}.
We thus have a unit cell with 16 atoms of which 4 are Si atoms, giving us a concentration of 25 at.\% $\approx$ 14.36 wt.\%.
Let's now do the same procedure as with bcc-Fe, we wish to have a k-mesh,
ecutwf and ecutrho that will work for both so we can later on ignore these
parameters. The unit cell has doubled in size so the k-mesh should be halved
let's see if we thus get convergence at (5,5,5). Starting with ecutwfc=64 and
ecutrho=782 as these, the minimal values for the Fe pseudopotential, are higher
than the ones in the Si pseudopotential. The results are shown in table \ref{table:DO3k-mesh}.
\begin{table}
\begin{tabular}{c|c|c}
	&k mesh&Hydrostatic Pressure (kbar)\\
	\hline
	1&(3,3,3)&-88\\
	2&(4,4,4)&-43\\
	3&(5,5,5)&-71\\
	4&(6,6,6)&-63\\
	5&(7,7,7)&-73\\
\end{tabular}
\caption{DO3-Fe$_3$Si k-mesh convergence}
\label{table:DO3k-mesh}
\end{table}\\
We "again" see convergence at (5,5,5). Now again keeping the cutoff charge density 12 times the cutoff
for the amount of wavefunctions we vary the charge density, as shown in table \ref{table:DO3ecutwfc}.
\begin{table}
\begin{tabular}{c|c|c}
	&ecutwfc&Hydrostatic Pressure (kbar)\\
	\hline
	1&14&-11640\\
	2&24&-1350\\
	3&34&-770\\
	4&44&-355\\
	5&54&-68\\
	6&64&-71\\
\end{tabular}
\caption{DO3-Fe$_3$Si ecutwfc convergence}
\label{table:DO3ecutwfc}
\end{table}\\
I.e we again see convergence at ecutwfc=54. Lastly we again search the multiplication factor, with results shown in table \ref{table:DO3factor}.
\begin{table}
\begin{tabular}{c|c|c}
	&factor&hydropressure\\
	\hline
	1&4&-12\\
	2&5&-61\\
	3&6&-50\\
	4&7&-63\\
	5&8&-68\\
\end{tabular}
\caption{DO3-Fe$_3$Si factor convergence}
\label{table:DO3factor}
\end{table}\\
We thus see convergence already at a factor of 5, so we'll take the highest factor of both, i.e 8. Our final 
values are thus ecutwfc=54, ecutrho at 432 and a k-mesh of (5,5,5).\\
Now, these values give stable and quite good results but whenever high precision results are necessary we'll do the calculation
with a k-mesh of (10,10,10) [(20,20,20) for Fe]  with ecutwf at 100 and ecutrho at 700. This "high precision" setting will be indicated by the $\dagger$ symbol.

\section{Energies of the end points}
In order to get a feeling for the accuracy of our results, but also to be able to calculate the stability of new hypothetical crystals, we need to calculate the energy of bcc-Fe and DO3-Fe$_3$Si, using the previously obtained basis sets.

We'll do a full geometry optimization by first using the
calculation="vc-relax" control parameter and bfgs cell and ion
dynamics with 0 pressure (0.5 kBar convergence threshold) for
both crystals.

\subsection{bcc-Fe}
Using the first method, for bcc-Fe we get a final total energy$^\dagger$ of -329.26 Ry for a unit cell containing 1 atom with a cell volume of 76.1642 a.u.$^3$
$\implies$ cell length of 4.238871 au, or a length of 
0.492942382 alat with a 2.999754 scale.
The calculated value for total magnetization is 2.13 $\mu$B/f.u. 

\subsection{DO3-Fe3Si}
For Fe$_3$Si we get a final total energy$^\dagger$ of -999.30084673 Ry for a unit cell containing 4 atoms with a cell volume of 296.49363 a.u.$^3$.
The calculated value for total magnetization is 5.07 $\mu$B/f.u.

\section{Sanity Check}
\begin{wrapfigure}{l}{0.1\textwidth}
  \begin{center}
	  \includegraphics[width=0.1\textwidth]{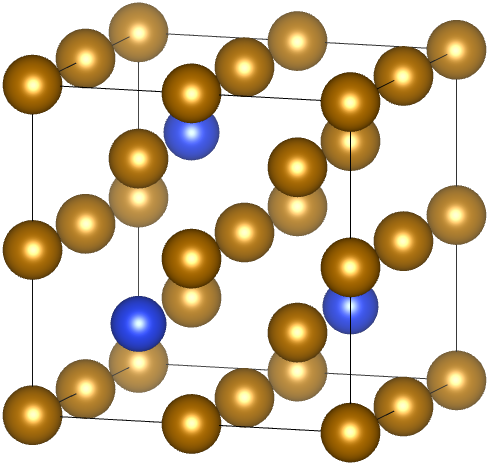}
  \end{center}
\end{wrapfigure}
We'll take the DO3-Fe$_3$Si and change 1 Silicon atom in the unit
cell to an Fe atom, for this we'll first have to remove the 
symmetry and define all the atom's coordinates ourselves.
The DO3-Fe$_3$Si crystal has F m -3 m as symmetry space group
(number 225). Looking at the crystallographic database
we see that the first Fe atom to consider which has relative atomic 
position (0.5 0.5 0.5) corresponding to Wyckoff Position 4b,
the second one has Wyckoff Position 8c and the Si atom has 4a.
I.e our unit cell has 4+8+4 = 16 atoms which is correct.
Making a cif file of a big unit cell can then be done quite
straightforwardly. We then check if the energy we get is indeed the
same. The energy we get from our Fe$_3$Si of a 16 atom unit cell
is -3997.2 Ry per 16 atom unit cell or -999.3 Ry per 4 which is the
same as we found earlier, this is thus a good input file to modify,
we'll now replace one Si atom with Iron as shown in the inline figure.
Doing this gives us a 18.75 at.\% $\equiv$ 10.40 wt.\% Si crystal.
After a 'vc-relax' calculation we get an energy of about -4314.944
Ry, this crystal can be composed of 3 Fe$_3$Si (containing 4 atoms)
and 4 Fe (containing 1 atom) unit cells giving a sum of energies of:
\begin{equation}
	3 \text{Fe}_3\text{Si} + 4 \text{Fe} = 3* E_{Fe_3Si} + 4*E_{Fe} \approx -4314.957 \; \mathrm{Ry}
\end{equation}
From this the formation energy (with respect to bcc-Fe and DO3-Fe$_3$Si) was calculated to be $E_{form} \approx + 0.011$ eV/atom. This means the energy of the modified Fe$_3$Si crystal is higher than the weighted sum of the bcc-Fe and DO3-Fe$_3$Si crystals; meaning it will decompose into those crystals.
\section{The grand search}
\begin{figure*}[ht]
\centering
    \begin{forest}
  [Grand Search
    [16
      [
        [
        1 [X]
        ]
        [
        2
          [
          X
          ]
        ]
        [
        3
          [
          X
          ]
        ]
      ]
    ]
    [
    32
     [
        [
        1
         [
         X
         ]
        ]
        [
        2
          [
          X
          ]
        ]
        [
        3
          [
          X
          ]
        ]
        [
        4
          [
          X
          ]
        ]
        [
        5
          [
          X
          ]
        ]
        [
        6
          [
          X
          ]
        ]
        [
        7
          [
          X
          ]
        ]
        [
        8
          [
          X
          ]
        ]
      ]
    ]
    [
    54
    [
    5
    [
    ?
    ]
    ]
    [
    6c
    [
    ?
    ]
    ]
    [
    6d
    [
    X
    ]
    ]
    [
    7
    [
    X
    ]
    ]
    ]
  ]
\end{forest}
\caption{Grand Search tree}
\label{fig:GStree}
\end{figure*}
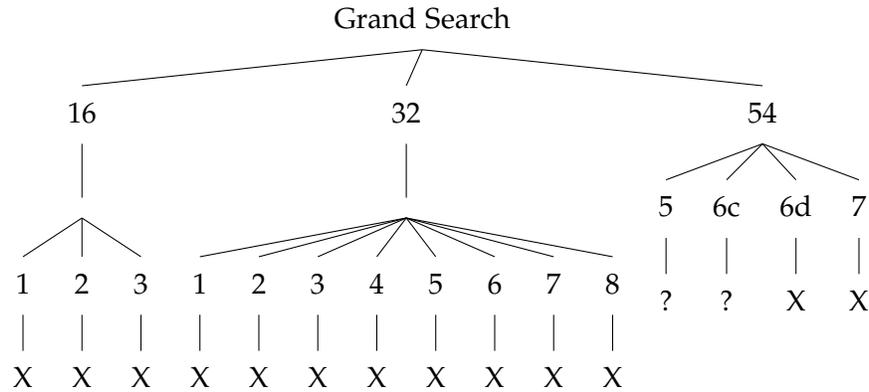

Having done all the previous calculations enables us to search for different (possibly) stable compositions of DO3-Fe$_3$Si and bcc-Fe. This was done by creating different supercells out of the original bcc-Fe and DO3-Fe$_3$Si unit cells. We made supercells consisting of 16, 32,54, 64 and 128 atoms and for each of those with different concentrations of silicon, however due to time 
and computational limits only the unit cells 16,32 and 54 were investigated.
As mentioned in the introduction, it would be optimal to find a stable crystal with 6.5 wt.\% Si.
For a unit cell of N atoms in total of which $\alpha$ are Si atoms,
the criterion is thus:
\begin{equation}
    0.065 \approx \frac{28.08\alpha}{28.08\alpha + 55.845(N-\alpha)}
\end{equation}
Or, 
\begin{equation}
\alpha = 0.1214642*N
\end{equation}
The unit cell choises are 16,32,54,64 and 128. The closest values to 
6.5wt.\% are given by:
\begin{table}[ht]
    \centering
    \begin{tabular}{ccc}
        N & $\alpha$ & wt.\%\\
        \hline
        16 & 2 & 6.7\\
        32 & 4 & 6.7\\
        54 & 7 & 6.9\\
        64 & 8 & 6.7\\
        128 & 16 & 6.7
    \end{tabular}
\end{table}\\
It's only at a unit cell of 256 that we deviate to the ratio 0.1210938 giving a closer wt.\% of $6.48$, we did not however
investigate this supercell as we don't have the needed computing power.
We did not only make supercells for 12.5 at.\% however, but also for 0-25 at.\% Si for every N (steps of $\frac{1}{N}$ at.\% Si).\\
In general, for a supercell (made from bcc-Fe and DO3-Fe$_3$Si unit cells) consisting of N atoms total, of which $\alpha$ are silicon, the formation energy is given by:
\begin{equation}
    \text{E}_{\text{form}}=\frac{E-(\alpha*\text{E}_{\text{DO3-Fe}_3\text{Si}}+(\text{N-}4\alpha)*\text{E}_{\text{bcc-Fe}})}{\text{N}}
\end{equation}
By calculating the formation energy we determine which compositions are energetically stable and which ones will decompose in bcc-Fe and DO3-Fe$_3$Si.
All calculations for the 16 and 32 supercells were vc-relax calculations, the tree structure representing our search is shown in figure \ref{fig:GStree}
The X'es indicate structures where the formation energy is positive (i.e it will
decompose) and the question marks '?' indicate structures where either there were technical
issues or the lattice should be further optimized as scf calculations were used instead of 
vc-relax calculations. For the base 54 unit cell the considered configurations were:

\begin{figure}[ht]
	\centering
	\begin{minipage}{.45\columnwidth}
		\centering
		\includegraphics[width=0.8\textwidth]{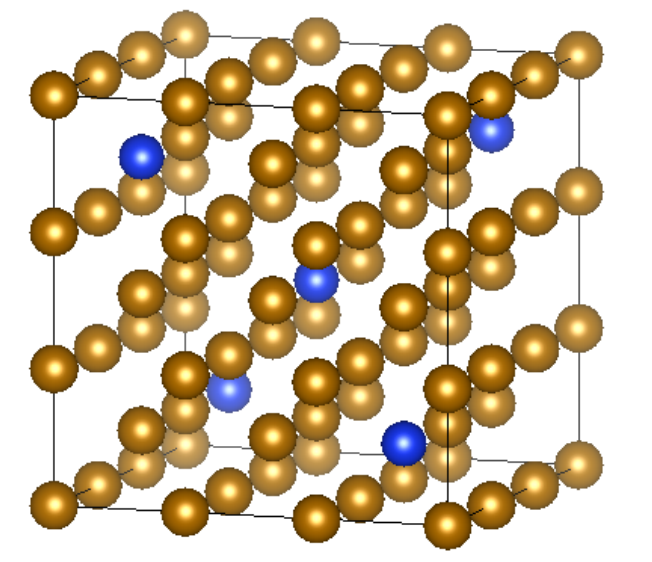}
            \caption{$\text{Fe}_{49}\text{Si}_5$}
            \label{fig:Fe49Si5}
	\end{minipage}%
  	\begin{minipage}{.45\columnwidth}
		\centering
            \includegraphics[width=0.8\textwidth]{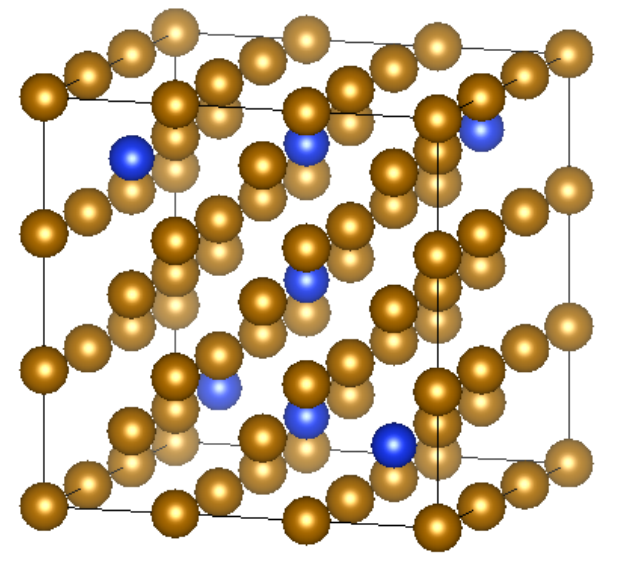}
            \caption{$\text{Fe}_{47}\text{Si}_7$}
            \label{fig:Fe47Si7}
	\end{minipage}
	\begin{minipage}{.45\columnwidth}
		\centering
		\includegraphics[width=0.8\textwidth]{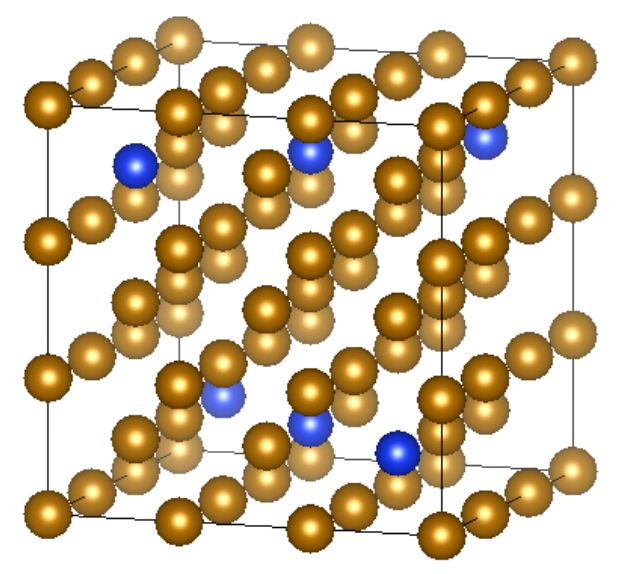}
            \caption{$\text{Fe}_{48}\text{Si}_6$ C}
            \label{fig:Fe48Si6C}
	\end{minipage}
 	\begin{minipage}{.45\columnwidth}
		\centering
            \includegraphics[width=0.8\textwidth]{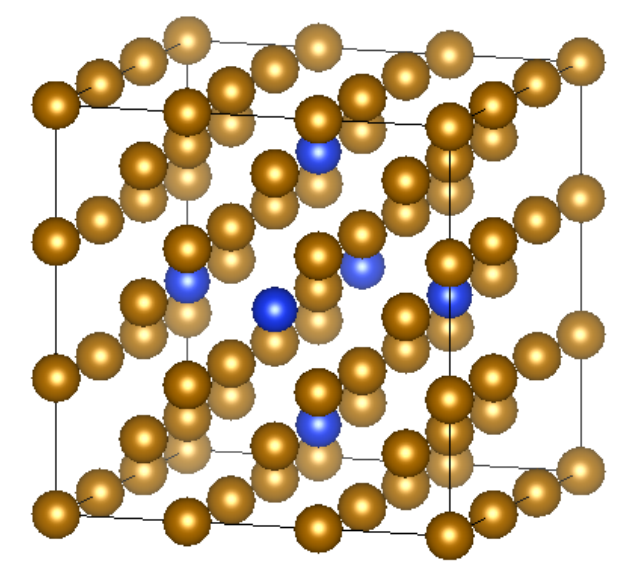}
            \caption{$\text{Fe}_{48}\text{Si}_6$\\  \hspace*{1.7cm}Diamond}
            \label{fig:Fe48Si6D}
	\end{minipage}
\end{figure}

From our computation, the most interesting crystals were $Fe_{49}Si{5}$ and $Fe_{48}Si{6}$ with a $E_{form}$ close to zero. Full optimization wasn't reached due to wall-time limit.

\section{Fractals}
\begin{wrapfigure}{l}{0.2\textwidth}
  \begin{center}
    \includegraphics[width=0.25\textwidth]{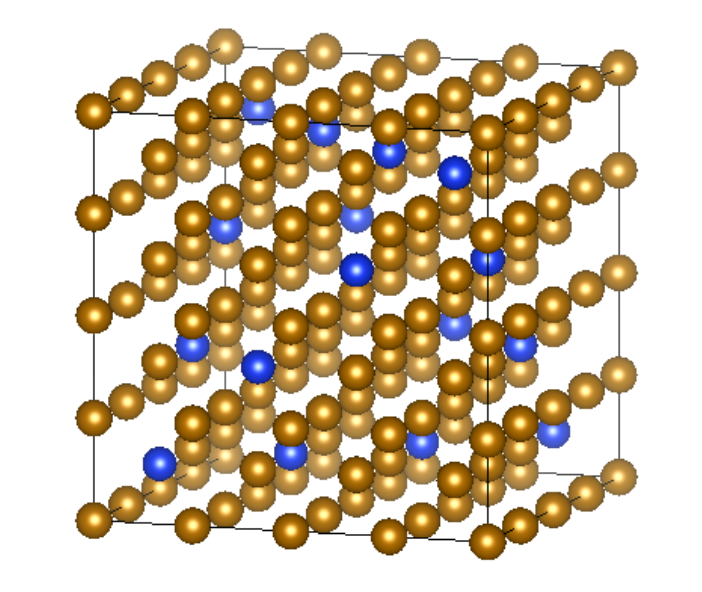}
    \caption{Unit cell of the proposed fractal}
    \label{fig:ProfFractal}
  \end{center}
\end{wrapfigure}
Professor S. Cottenier suggested a fractal (self-similar) structure of which the unit cell is shown in figure \ref{fig:ProfFractal}. This is a unit cell built up out of 4 DO3-Fe$_3$Si cubes and 4 Fe cubes in such a way to make the DO3 cubes form a tetrahedron similar to the tetrahedron of Si atoms in DO3-Fe$_3$Si. That structure is made of 16 atoms of Si and 112 atoms of Fe.
The total formation energy of such a unit cell is given by: $E_f = 16*E_{DO3-Fe_3Si}+(128-4*16)*E_{bcc-Fe} = -37062.5 Ry$ With a wt.\% of 6.7\%.\\
We tried to compute the energy for such a cell but couldn't obtain a significant result. The best results and parameters are shown in table \ref{fractal_table}.
The computations were made on the VSC and in mode \emph{scf}.
Many computations were needed to obtain both those results and the right parameters (K, cut off energy, number of core and wall time).
The convergence problem could be tackled by doing a vc-relax calculation with bigger k-mesh and basis set size, thus obtaining both cell and position parameters whom would be closer to reality. Due to time limits however\footnote{And issues with the VSC in Ghent during the start of december} this
couldn't be done.

\section{Conclusion}
Even though significant results weren't found, either as there wasn't enough time or due 
to technical reasons (e.g the VSC in Ghent being under maintenance). This paper may
lay the ground work for more research to come as anyone who stumbles on this paper knows which
parameters to use and can go onto the project page where the cell files are located, running 
one of the calculations her/himself and maybe obtaining the yet sought after answer.
\begin{table*}[ht]
\centering
\resizebox{\textwidth}{!}{%
\begin{tabular}{|c|c|c|c|c|c|c|c|}
\hline
Trial & K mesh & ecutwfc & ecutrho & E {[}Ry{]} & E accuracy {[}Ry{]} & E$_{\text{form}}$ & Error \\ \hline
1 & 1 & 27 & 216 & -36921.4 & 3837.5 & 1.103 & Did not converge after 100 iterations \\ \hline
2 & 2 & 27 & 216 & -37010.4 & 2.6 & 0.407 & Hit wall time after 27 iterations \\ \hline
3 & 2 & 54 & 432 & -37042.0 & 546.9 & 0.160 & Did not converge after 100 iterations \\ \hline
4 & 3 & 27 & 216 & -36923.2 & 2254.3 & 1.088 & Did not converge after 100 iterations \\ \hline
5 & 3 & 54 & 432 & -36971.6 & 1717.7 & 1.683 & Did not converge after 100 iterations \\ \hline
6 & 3 & 60 & 600 & -36979.4 & 2064.3 & 1.540 & Did not converge after 100 iterations \\ \hline
\end{tabular}%
}
\caption{Fractal computation results}
\label{fractal_table}
\end{table*}

\bibliography{sources}
\bibliographystyle{plain}


\end{document}